# The interpretation of the Einstein-Rupp experiments and their influence on the history of quantum mechanics[a]

Jeroen van Dongen[b]

*Institute for History and Foundations of Science*
*Utrecht University*
*PO Box 80.000, 3508 TA Utrecht, the Netherlands*
*&*
*Einstein Papers Project, California Institute of Technology, Pasadena, CA 91125 USA*

**Abstract**

The Einstein-Rupp experiments were proposed in 1926 by Albert Einstein to study the wave versus particle nature of light. Einstein presented a theoretical analysis of these experiments to the Berlin Academy together with results of Emil Rupp, who claimed to have successfully carried them out. However, as the preceding paper has shown, this success was the result of scientific fraud. This paper will argue, after exploring their interpretation, that the experiments were a relevant part of the background to such celebrated contributions to quantum mechanics as Born's statistical interpretation of the wave function and Heisenberg's uncertainty principle. Yet, the Einstein-Rupp experiments have hardly received attention in the history of quantum mechanics literature. In part, this is a consequence of self-censorship in the physics community, enforced in the wake of the Rupp affair. Self-censorship among historians of physics may however also have played a role.

In the spring of 1926, Albert Einstein proposed to Emil Rupp to do two experiments that were to probe the wave versus particle nature of light: the so-called 'Wire Grid Experiment' and the 'Rotated Mirror Experiment' (*Spiegeldrehversuch*). In both experiments, the interference properties of light emitted by canal ray sources were to be explored to reveal if light was emitted in a process that was extended in time, as was to be expected on the basis of its classical description as a wave, or whether it was emitted instantaneously. The foregoing paper took up these experiments in detail, including strong evidence that Rupp's results were in fact fraudulent.[*] The present paper raises two related questions: First, how did Einstein accommodate the conflicting notions of wave and particle in the context of the experiments? Second, how might these experiments have influenced contemporary developments? In both respects, even without taking up all possible ramifications, the episodes discussed here suggest that the Einstein-Rupp experiments played a relevant, perhaps even positive role in the construction of quantum mechanics.

---

[*] Jeroen van Dongen, "Emil Rupp, Albert Einstein and the canal ray experiments on wave-particle duality: Scientific fraud and theoretical bias," *Historical Studies in the Physical and Biological Sciences* **37** Suppl. (2007) 73-120.

**Interpretation: Einstein on waves, particles, and ghost fields**

Einstein's interest in the canal ray experiments went back to his desire to test wave and particle pictures of light; to see "how much of either is correct."[1] As the preceding paper shows, he initially expected a clear confirmation of the particulate, instantaneous emission picture in the canal ray experiments. On the other hand, despite that Einstein's theoretical prejudices heavily determined his interactions with Rupp, once he realized that the latter had already "unknowingly"[2] confirmed the classical wave picture, he gradually reshaped his views. Indeed, Einstein soon began to expect further confirmations of the wave picture and later claimed that Rupp's experiments had given the classical result.

One important role of the Einstein-Rupp experiments is thus easily identified: they maintained a wave-picture of light at a crucial moment during the genesis of the quantum theory[3]—just as experiments by in particular Arthur Compton had confirmed its particulate aspects.[4] Einstein of course had already early on pointed out that light exhibited both wave and particle properties, for instance in his study of the energy fluctuations in black body radiation.[5] Given these contexts, and Einstein's initial expectations and gradual turn-around, one should expect that he had a dual wave-particle picture of light when the canal ray experiments were under discussion in the spring of 1926. But what could the details of that dual picture have looked like?

Here we must turn to Einstein's ideas on the "ghost field" description of light: a probabilistic and dual interpretation of light attributed to Einstein[6] but never explicitly published by him as such. As John Stachel has also suggested, to get a sense of these ideas we must go back to 1921[7] when traces of this interpretation appear in Einstein's correspondence. In particular, Hendrik Antoon Lorentz wrote to Einstein in November of 1921—in the context of Einstein's proposed, but flawed, canal ray experiment of that year—and outlined in his letter a probabilistic interpretation of the light wave that he attributed to Einstein. According to Lorentz's reconstruction of Einstein's thoughts, the latter held that:

> In light emission, two things are emitted. There is namely: 1. An interference radiation, that occurs according to the normal laws of optics, but still carries no energy. One can for example imagine that this radiation exists in normal electromagnetic waves but with vanishingly small amplitudes. As a consequence they cannot themselves be observed; they only serve to prepare the way for the radiation of energy. It is like a dead pattern, that is first brought to life by the energy radiation. 2. The energy radiation. This consists of indivisible quanta of magnitude $h\nu$. Their path is prescribed by the (vanishingly small) energy flux in the interference radiation, and they can never reach places where this flux is zero (dark interference bands).

In an individual act of radiation the full interference radiation arises, but only a single quantum is radiated, which therefore can only reach one place on a screen placed in the radiation. However, this elementary act is repeated innumerably many times, with as good as identical interference radiation (the same pattern). The different quanta now distribute themselves statistically over the pattern, in the sense that the average number of them at each point of the screen is proportional to the intensity of the interference radiation reaching that point. In this way the observed interference phenomena arise, corresponding to the classical results.[8]

Lorentz continued by outlining a suggestive idea of his own:

[W]e do now not need to conclude that, in the case that an interference phenomenon with a phase difference of *N* (for example $10^6$) wavelengths is observed, the quantum has to stretch itself in the direction of propagation over *N* wavelengths. It can very well be quite small. When in an elementary emission event (with an energy quantum) a train of *N* waves (interference radiation) is emitted, one can raise the question where in that train the single quantum is; up front or in the back, or [it] could take up roughly all positions in between, and when often repeated also really does. One could conclude something about this from observations of the visibility of interference fringes at various path differences. Namely, the following is to be taken into consideration: Let us assume that a screen *S* is hit by the two wave trains 1 and 2 (that originated at the same emission event), with front and rear wavefronts *a* and *b*, or *c* and *d* respectively [see figure 1]. A light quantum can only make the

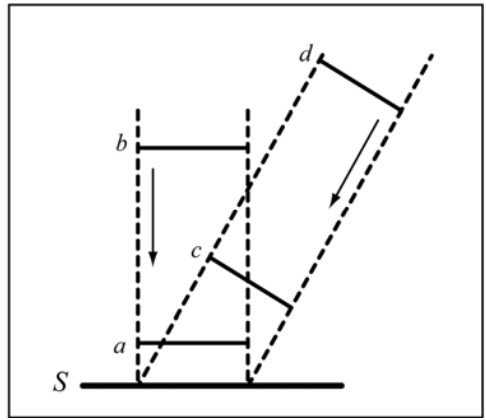

Figure 1: Copied after Lorentz's figure in his letter to Einstein of 13 November 1921, EA 16 544.

interference visible if, at the very moment that it lights up the screen, on the latter there is already interference in the interference field. That is, if both rays of the interference field overlap. If the screen is reached by 2 somewhat later than by 1, then the light quanta that are very much to the front in 1 or to the back in 2 can not produce any sharp fringes, etc.[9]

This idea of Lorentz essentially resurfaced in Einstein's Wire Grid Experiment, where the cutting up into the "two wave trains 1 and 2" of the interference field would occur because of the grid. If in the Wire Grid Experiment a variability in the visibility of the interference with the path difference would be observed, it could easily be understood in terms of Lorentz's interpretation—based on Einstein's ghost field—given above: the production of the interference field would take an extended lapse of time and its fringe pattern would give a probability distribution according to which the individual quanta would arrange themselves on the screen. In the minima of the visibility of fringes, the cut up wave trains of the interference field do not overlap and no pattern can form. If no minima in the visibility were observed—as Einstein initially expected—then one could conclude that the interference field might somehow be instantaneously emitted or transmitted through the grid. However, it should be pointed out that Einstein, in the case of such an outcome, originally only expressed the expectation that the "sine-like character of the wave field" would not be "conditioned by the emitting atom or electron," but by "conditions imposed by specific laws of the space-time continuum."[10] He did not further elaborate on these presumed laws, and neither on how they would condition the wave field, but only stated that in the case of a negative outcome of the Wire Grid Experiment one could conclude that interference had nothing to do with any periodicity of the radiating atom. Rupp's results of course contradicted such conclusions.

The above congruence between Lorentz's idea and the Grid Experiment strongly suggests that some form of the ghost field interpretation was on Einstein's mind when he proposed the experiment in 1926. One can also easily see how it would apply to the *Spiegeldrehversuch*, though there is again no concrete evidence that Einstein in fact did so. Unfortunately, it is difficult to reconstruct his full interpretation on the basis of the documentary evidence.

However, the inferences that Einstein drew in 1926 on the basis of Rupp's claims do point in this direction, as far as they can be reconstructed from his correspondence with the latter. After Rupp had submitted his manuscript, Einstein reviewed it and came across a statement that he disagreed with—apparently, Rupp believed that one could conclude from his experiments that the atom gradually passes from an excited to a non-excited state. Einstein urged Rupp to change this passage and expressed that in his view:

> One must distinguish between the production of the interference field (A) and the energy emission (B). The event-like nature of (B) is certain. Your experiments have proven that (A) is a process that is extended in time. Whether (A) takes place while the atom is in its excited state, that is, contains the full $h\nu$, is indeed not certain.[11]

Rupp did not reply timely and Einstein decided to make the changes himself. In his next letter, he again emphasized that "it is today really rather certain that the undulatory and the energy properties must be clearly separated, as only the latter have an instantaneous character."[12]

The separation of the "interference field" and the energy properties of light are in full agreement with the probabilistic ghost field interpretation as encountered in Lorentz's letter of 1921. Yet, there is no mention of a probability distribution; on the basis of these sources alone one can assert no more than that Einstein made the plain observation that an interference field is emitted along with the light quanta, and that the emission of the interference field takes an extended lapse of time. In his Academy publication however, Einstein would not even go that far and did not mention the interference field; he only concluded that the classical extended-in-time predictions were correct (although he did hint in a footnote that "one is not allowed to conclude that the quantum process of emission, that in terms of energy is completely determined by location, time, direction and energy [sic], is also *geometrically* determined by these quantities.")[13]

Einstein's reserved attitude regarding the details of his understanding of light's duality is perhaps best illustrated, finally, in his lecture at Berlin University of 23 February 1927.[14] In this seminar on "theory and experiment on the question of the origin of light," he again left the question open. After first outlining the dilemma—wave or particle—he spoke of "detailed experiments, carried out by Dr. Rupp" that had confirmed that emission is a process that takes an extended period of time. Einstein here did emphasize the need to sharply separate between the "energy" and "geometric" properties of light, but he did not discuss a probabilistic ghost field interpretation. Instead, he concluded that "what nature asks of us, is not a quantum theory or wave theory, but nature asks of us a synthesis of both views that so far has exceeded the intellectual powers of physicists."

**Possible ramifications: Born and Heisenberg**

Historians of physics have already pointed to the close relation between Einstein's ghost field interpretation, as contained in Lorentz's 1921 letter, and the Born interpretation of the wave function $\psi$.[15] In his 1954 Nobel lecture, Born spoke of the key developments that had led him to his result. He in particular stated that:

> [A]n idea of Einstein gave me the lead. He had tried to make the duality of particles—light quanta or photons—and waves comprehensible by interpreting the square of the optical wave amplitudes as probability density for the occurrence of photons. This concept could at once be carried over to the $\psi$-function: $|\psi|^2$ ought to represent the probability density for electrons (or other particles).[16]

Einstein's influence is evident in Born's original publications too: "[...] I tie in with a remark by Einstein on the relation between wave field and light quanta; he said more or less that the waves are only there to show the way to the corpuscular light quanta, and he spoke in this sense of a 'ghost field.' This determines the probability that a light quantum, the carrier of energy and momentum, takes a particular direction; the field itself does not contain any energy or momentum."[17] Born further suggested to carry this idea over from the electromagnetic field to the Schrödinger wave field, and interpret this as a "ghost field" too. He then went on to formulate his interpretation in the context of an electron scattering off an atom.

The two papers in which Born made this step were submitted on 25 June and 21 July, 1926, just when Einstein was concluding his collaboration with Rupp and the latter had begun drawing up his Academy paper.[18] Born and Einstein frequently interacted, so it is very well possible that they discussed the Einstein-Rupp experiments in the spring of 1926.[19] Even if Born and Einstein had not actually discussed the experiments, Einstein's March 1926 publication in *Naturwissenschaften*[20] already prominently drew attention to the Wire Grid Experiment. Clearly, these coincidences do not warrant one to state as a fact that Born had the Einstein-Rupp experiments on his mind when he formulated his interpretation—however, they do make it plausible that the experiments may have played a role.

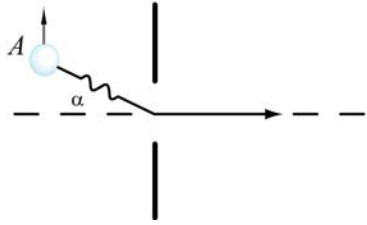

Figure 2: Atom *A* that emits light passes behind a slit. The light is diffracted at an angle $\alpha$ with the normal.

Werner Heisenberg's 1929 Chicago lectures[21] illustrate the important role of the Einstein-Rupp experiments even more directly. Heisenberg used the Wire Grid Experiment to show how one might suspect a contradiction between the wave and particle picture in the case of an atom moving with velocity *v* past a slit of width *d*. Namely, an emitted light wave would be cut up by the slit and therefore have a spread in its frequency of the order of $\Delta \nu \sim v/d$. However, according to the light quantum theory, the emitted light is strictly monochromatic with its energy given by $h\nu$. There is no contradiction however, if one takes into account the fact that the quanta undergo diffraction at the slit, an idea that Heisenberg credited to Bohr. Quanta emitted at an angle $\alpha$ with the normal also reach a point on that normal behind the slit, with $\alpha$ of the order of $\sin \alpha \sim \lambda/d$ (see figure 2). These quanta have undergone a Doppler shift: $\Delta \nu \sim \sin \alpha \times \frac{v}{c} \times \nu$. From this then again followed $\Delta \nu \sim v/d$. The particle picture is thus consistent with the wave picture and Heisenberg concluded that in this experiment "strict validity of the energy law for particles is in agreement with the demands of classical optics."[22]

Traces of the alluded to discussion between Heisenberg and Bohr on the Einstein-Rupp experiments can be found

in the Einstein Archive. Shortly before the appearance of Heisenberg's article that contained the uncertainty relations,[23] Bohr wrote a letter to Einstein in which he advertised Heisenberg's results; he did so in the context of Einstein and Rupp's Grid Experiment ("I would like to add some comments that connect to the problem that you have recently discussed in the Proceedings of the Berlin Academy.")[24] After first arguing that the concept of a finite wave train was in good agreement with the uncertainty principle for quanta,[25] Bohr discussed essentially the same argument as Heisenberg would later present in his Chicago lectures. He added to Einstein that "as you hinted at in your footnote any 'light quantum description' can never explicitly account for the geometrical relations of the 'radiation trajectory.'"[26] With Heisenberg's new results, energy conservation of particles and wave optics could be brought into agreement in the Wire Grid Experiment, as "the two sides of the problem never surface at the same time according to the nature of the description."[27] Heisenberg in his paper on the uncertainty relations stated that his ideas originated partly in "Einstein's discussions on the relation between wave field and light quanta."[28] He may not have been thinking of Einstein's ghost fields here—as the wave-particle complementarity hinted at by Bohr is of course a different concept than a ghost field interpretation, since in the latter the particles and waves are present simultaneously—yet the context of the Einstein-Rupp experiments appears to be relevant again.

As before, the sources do not spell out in full detail what influence Einstein's theoretical paper and Rupp's experimental publication exerted on discussions between Heisenberg and Bohr. It seems however safe to conclude that these experiments were involved in communicating the uncertainty relations, and that they likely had a part in Bohr and Heisenberg's development of key conceptual elements of the quantum theory.

**Afterwards**

Bohr did not mention the Einstein-Rupp experiments in his 1949 review of his exchanges with Einstein on the foundations of quantum theory.[29] Indeed, despite their obvious place in Einstein's oeuvre and despite their widespread contemporary reception, the experiments are hardly discussed in the Einstein literature.[30] Similarly and perhaps surprisingly, this is likely the first occasion that the Einstein-Rupp experiments have been pointed out as relevant context for Born's references to Einstein. In a *Science* paper, Abraham Pais, the noted Einstein biographer, also emphasized the role of Einstein's thoughts on the "ghost field" as an inspiration for Born. Yet, he did not mention the Einstein-Rupp experiments in his account of Born's creative moment, nor did he take them up in his

Einstein biography.[31] Scholarship on Heisenberg has also not yet addressed the point.[32]

It seems as if the German Physical Society's decision not to allow citations to Rupp's fraudulent work has tacitly been observed in the historical literature. One hardly finds any mention of Rupp—let alone of the fraud that he committed in the canal ray experiments—in historical studies of either quantum theory or of Einstein.[33] This may be due to a genuine failure to notice Rupp's role, precisely since references to his work became scarce, or perhaps to a desire to maintain an untainted image of Einstein or a tidy account of the transition from classical to quantum theory. Yet, although Rupp committed fraud, it appears that this did not directly hamper the progression to quantum mechanics. He claimed to have confirmed Einstein's theoretical intuition and this (revised) intuition in the end turned out to be in line with the fully developed quantum theory. In that theory, the Copenhagen doctrine of complementarity entails that the experimental environment dictates the conceptual interpretation of the experiment. In the case of the Einstein-Rupp experiments, that implies that one should expect a confirmation of the wave picture for radiation, just as Einstein in the end predicted and Rupp claimed to have observed.

---

[a] To appear in: *Historical Studies in the Physical and Biological Sciences*, *37* Suppl. (2007), 121-131.

[b] J.vanDongen@phys.uu.nl. I would like to thank Cathryn Carson, Dennis Dieks, Michel Janssen, A.J. Kox and Jos Uffink for insightful discussions and helpful suggestions, and the Einstein Archive at Hebrew University and Princeton University Press for permission to use excerpts from Einstein's correspondence in this article. I gratefully acknowledge support by a Veni-grant of the Netherlands Organization for Scientific Research (NWO). The following abbreviations are used: EA, Albert Einstein Archives, Hebrew University, Jerusalem; *ZfP*, *Zeitschrift für Physik*.

[1] Albert Einstein to Max von Laue, 29 Aug 1936, EA 16 113.

[2] Emil Rupp to Einstein, 21 Aug 1926, EA 20 402.

[3] See also Jagdish Mehra and Helmuth Rechenberg, *The historical development of quantum mechanics, volume 6: The completion of quantum mechanics, 1926-1941* (New York, Berlin, Heidelberg, 2001), 235-236.

[30] As just one example, Albrecht Fölsing's biography does not mention Emil Rupp, whereas in an earlier book on scientific fraud, he did in fact briefly discuss Rupp's canal ray experiments: Albrecht Fölsing, *Albert Einstein: Eine Biografie* (Frankfurt a. M., 1993); Albrecht Fölsing, *Der Mogelfaktor. Die Wissenschaftler und die Wahrheit* (Hamburg, 1984).

[31] Pais (ref. 15); Pais (ref. 4).

[32] See e.g. Kristian Camilleri, "Heisenberg and the wave-particle duality," *Studies in history and philosophy of modern physics*, *37* (2006), 298-315.

[33] The Einstein-Rupp experiments are e.g. not mentioned in Max Jammer, *The conceptual development of quantum mechanics* (New York, 1966). Helge Kragh noted that in the multi-volume history of quantum mechanics by Jagdish Mehra and Helmut Rechenberg, Rupp's fraud was barely pointed out; Helge Kragh, "Book review," *Foundations of physics*, *32* (2002), 187-189, review of: Jagdish Mehra and Helmuth Rechenberg, 2001 (ref. 3). Mehra and Rechenberg do discuss the Einstein-Rupp experiments (pp. 235-236), but they only qualify later work of Rupp as controversial (p. 379). A.P. French, "The strange case of Emil Rupp," *Physics in perspective*, *1* (1999), 3-21, clearly is the exception to the above observation, as well as some of the secondary literature cited in it.